\documentclass[conference]{IEEEtran}
\IEEEoverridecommandlockouts

\usepackage{epsfig,psfrag}

\usepackage{amssymb}
\usepackage[cmex10]{amsmath}
\usepackage{stmaryrd}
\newtheorem{theorem}{Theorem}[section]

\newtheorem{corollary}[theorem]{Corollary}

\begin{document}

\title{On the Queueing Behavior of Random Codes\\
over a Gilbert-Elliot Erasure Channel}

\author{\IEEEauthorblockN{Parimal Parag,
Jean-Francois Chamberland,
Henry D. Pfister,
Krishna R. Narayanan}
\IEEEauthorblockA{Department of Electrical and Computer Engineering, Texas A{\&}M University}
\vspace{-5mm}
\thanks{This material is based upon work supported, in part, by the National Science Foundation (NSF) under Grant No.~0830696, by the Texas Norman Hackerman Advanced Research Program under Grant No.~000512-0168-2007, and by Qatar Telecom (Qtel).
Any opinions, findings, conclusions, and recommendations expressed in this material are those of the authors and do not necessarily reflect NSF's, THECB's or Qtel's views.}
}

\maketitle

\begin{abstract}
This paper considers the queueing performance of a system that transmits coded data over a time-varying erasure channel.
In our model, the queue length and channel state together form a Markov chain that depends on the system parameters.
This gives a framework that allows a rigorous analysis of the queue as a function of the code rate.
Most prior work in this area either ignores block-length (e.g., fluid models) or assumes error-free communication using finite codes.
This work enables one to determine when such assumptions provide good, or bad, approximations of true behavior.
Moreover, it offers a new approach to optimize parameters and evaluate performance.
This can be valuable for delay-sensitive systems that employ short block lengths.
\end{abstract}

\section{Introduction}

Forward error-correcting codes have played an instrumental role in the many successes of digital communications over the past decades~\cite{Costello1998tit}.
The fact that it is possible to transmit digital information reliably at a positive rate over an unknown noisy channel is now universally acknowledged~\cite{Lapidoth1998tit}.
The main cost of improving reliability is the use of increasingly long codewords~\cite{Gallager0471290483}.
%
One situation where the valuable lessons of classical coding theory may not apply directly is the general area of delay-constrained communications.
If system specifications dictate that almost all information bits should be made available at the destination shortly after they arrived at the transmitter, it may not be possible to aggregate a large number of them before encoding and transmission.
In some cases, stringent delay requirements will force a system designer to resort to short block codes or short constraint-length convolutional codes.

From a coding perspective, using short codewords on channels with memory creates two impediments.
First, decoders are designed to correct the most-likely error patterns and the probability of seeing atypical error patterns cannot be neglected for short block lengths.
Second, if the coherence time of the channel is longer than a codeword transmission interval, then optimal code rate may depend heavily on the channel state, which is unknown to the transmitter.
Together, these factors impair the rapid transmission of information.

Coding performance as a function of block-length and code-rate has been assessed in the information theory literature using the reliability function~\cite{Gallager0471290483}.
This criterion focuses on the exponential rate at which the error probability decays with block length, known as the error exponent, as a function of information rate.
The concept of a reliability function can also be extended to variable-length codes in the presence of feedback~\cite{Burnashev1976}.
More recently, consideration has been given to the reliability function for bits with fixed delay, as opposed to coded blocks, in the presence of feedback~\cite{Sahai2008tit}.

While remarkable, these results remain asymptotic in nature and do not necessarily capture overall system behavior adequately.
For delay-sensitive applications and short codewords, three interrelated effects come into play.
The probability of decoding failure for every codeword is not negligible.
Packet retransmissions lead to queue buildups at the source and, thereby, induce longer latencies.
Channel correlation over time introduces dependencies among successive decoding attempts, which further perturb queueing behavior and end-to-end delay.
This is especially true when decoding failures are likely to occur in sequence~\cite{Liu2007tit}.
Thus, a queueing analysis is necessary when considering the behavior of communication systems subject to very stringent delay requirements.

For delay-sensitive systems with short codewords, the natural tradeoff between code-rate and probability of decoding failure is hard to characterize~\cite{Ephremides1998tit}.
In a non-asymptotic regime where information is queued at the source, transmitting data at a rate slightly below Shannon capacity may lead to poor performance.
Recent results in the literature hint at the fact that, for delay-constrained communication, optimal code-rate selection depends heavily on block-length and channel correlation~\cite{Wu2003twcom,Ying2008tit}.
These findings are especially important for real-time traffic and live interactive sessions, as these applications are sensitive to latency and require the use of short codewords.

Guidelines for code-rate selection in the context of delay-sensitive traffic were previously obtained for an erasure channel with memory~\cite{ParimalITW2010}.
The approach favored therein, which permits a complete characterization of queueing behavior, consists in building a Markov model for the evolution of the system.
Crucial assumptions that facilitate analysis can be summarized as follows:
the packet arrival process at the source is Bernoulli, the packet lengths are i.i.d. geometric, the error protection uses random codes, and the channel evolution is governed by a Markov chain.

In this article, we adopt a similar formulation and extend results that were obtained for the correlated erasure case to a more encompassing Gilbert-Elliot framework.
This latter class of erasure channels is common to the literature on channels with memory, and subsumes earlier work based on similar concepts.
We also present an in-depth analysis of system performance using different criteria that reflect the needs of various contemporary applications.
This research is significant because it offers a new perspective on the selection of code-rate and block-length for delay-sensitive systems and provides a rigorous investigation into the effects of time-correlation on the queued performance of real-time wireless connections.


\section{Channel Abstraction and Coding}
\label{section:AbstractModel}

\begin{figure}[t]
\begin{center}
\begin{psfrags}
\psfrag{g}[c]{$g$}
\psfrag{b}[c]{$b$}
\psfrag{Pg}[c]{$\varepsilon_g$}
\psfrag{Pb}[c]{$\varepsilon_b$}
\psfrag{u}[l]{$\alpha$}
\psfrag{d}[l]{$\beta$}
\psfrag{0}[c]{$0$}
\psfrag{1}[c]{$1$}
\psfrag{e}[c]{$e$}
\epsfig{file=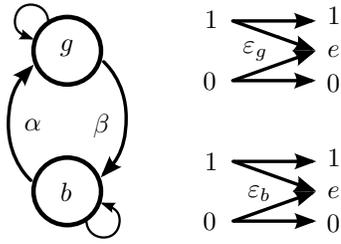,width=4.455cm}
\end{psfrags}
\caption{A Gilbert-Elliot bit erasure channel is employed to model the operation of a communication link with memory.
This model captures both the uncertainty associated with transmitting bits over a noisy channel and correlation over time typical of several communication channels.}
\label{figure:GilbertElliotChannel}
\end{center}
\end{figure}

Throughout, we assume that coded bits are sent from the transmitter to the destination over a Gilbert-Elliot erasure channel.
This channel can be in one of two states: a \emph{good} state $g$ in which every bit is erased with probability $\varepsilon_g$ and a \emph{bad} state $b$ in which every bit is erased with probability $\varepsilon_b$, independently of other bits.
Our naming scheme implies $\varepsilon_b \geq \varepsilon_g$.
Transitions between channel states occur according to a Markov process.
The probability of transitioning to state $g$ given that the Markov chain is currently in state $b$ is denoted by $\alpha$.
The likelihood of the reverse transition from $g$ to $b$ is symbolized by $\beta$.
Under alphabetical state ordering, the parameters of this Markov chain can be expressed in the form of a probability transition matrix,
\begin{equation} \label{equation:StateTransition}
\mathbf{P} = \left[ \begin{array}{cc} 1 - \alpha & \alpha \\
\beta & 1 - \beta \end{array} \right] .
\end{equation}
A graphical interpretation of the communication channel under consideration appears in Fig.~\ref{figure:GilbertElliotChannel}.

The state of the channel at time $n$ is a random variable, which we denote by $C_n$.
Moreover, the succession of states over time, $\{ C_n : n \in \mathbb{N} \}$, forms a Markov process.
Finding the conditional probability $\Pr (C_{n+1} = d | C_n = c)$ amounts to selecting an entry in $\mathbf{P}$.
Likewise, $\Pr ( C_{n+N} = d | C_n = c )$ can be obtained by locating the corresponding entry in $\mathbf{P}^N$, the $N$th power of $\mathbf{P}$.
We note that this Markov chain converges to its stationary distribution at an exponential rate that depends on the second eigenvalue of $\mathbf{P}$ (i.e.\ $1 - \alpha - \beta$).

In our analysis, a packet of length $L$ is sectioned into $M$ data segments each containing $K$ information bits.
Packing loss is treated implicitly since the last data segment of each packet is zero padded to $K$ bits.
Every segment is encoded separately into a codeword of length $N$, which is subsequently stored in the queue for eventual transmission over the Gilbert-Elliot erasure channel.
Decoding failures are handled through immediate retransmission of the missing data.

\subsection{Distribution of Erasures}
\label{section:ErasureDistribution}

A quantity that is of fundamental importance in our analysis is the conditional probability of decoding failure at the destination.
An intermediary step in identifying this probability is to derive an expression for $E$, the number of erasures within a codeword of length $N$.
This, in turn, depends on the number of visits to each state within $N$ consecutive realizations of the channel.
More specifically, we are interested in conditional probabilities of the form
\begin{equation} \label{equation:NumberErasures}
\Pr (E = e, C_{N+1} = d | C_1 = c) ,
\end{equation}
where $e \in \mathbb{N}_0$ and $c, d \in \{ b, g \}$.
The generating function for these conditional probabilities is based on generalizing the entries of $\mathbf{P}$ to the vector space of real polynomials in $x$ with
\begin{equation*}
\mathbf{P}_x = \left[ \begin{array}{cc} (1 - \alpha) (1 - \varepsilon_b + \varepsilon_b x) & \alpha (1 - \varepsilon_b + \varepsilon_b x) \\
\beta (1 - \varepsilon_g + \varepsilon_g x)
& (1 - \beta) (1 - \varepsilon_g + \varepsilon_g x) \end{array} \right].
\end{equation*}
Let $\llbracket x^j \rrbracket$ be the operator which maps a polynomial in $x$ to the coefficient of $x^j$.
Then, the conditional probability \eqref{equation:NumberErasures} is given, in terms of the $N$th power of $\mathbf{P}_x$, by
\[ \Pr (E = e, C_{N+1} = d | C_1 = c) = \llbracket x^e \rrbracket \left[ \mathbf{P}_x^N \right]_{c,d}. \]
It is worth mentioning that one can employ this method or alternative combinatorial means to obtain closed-form expressions for the desired conditional probabilities~\cite{Wilhelmsson-com99,ParimalITW2010}.


\subsection{Probability of Decoding Failure}
\label{section:BlockFailure}

During every transmission, a segment of $K$ information bits is encoded using a code defined by a random parity-check matrix $\mathbf{H}$ of size $(N\!-\!K) \times K$, where each matrix entry is selected independently and uniformly from $\{ 0, 1 \}$.
Maximum likelihood decoding is used at the destination.

Random coding has the benefit that the probability of decoding failure depends only on the number of erasures and not on the locations of the erasures.
Consequently, the decoding failure probability is a function of the number of erasures $E$ in the block.
Once the value of $E$ is known, we can derive the desired probability as follows.
Conditioned on $E=e$, decoding at the destination will succeed if and only if the submatrix of $\mathbf{H}$ formed by choosing the $e$ erased columns has rank $e$~\cite{Richardson0521852293}.
Furthermore, the probability that a random $e \times p$ matrix over $\mathbb{F}_2$, where $p = N \!-\! K$ stands for the number of parity bits, has rank $e$ is equal to
$\prod_{i=0}^{e-1} \left( 1- 2^{i-p} \right)$.
Thus, given $e$ erasures within a codeword of length $N$, the probability of decoding failure can be written as
\begin{equation*}
P_{\mathrm{f}}(N\!-\!K,e) \triangleq 1 - \prod_{i=0}^{e-1} \left( 1- 2^{i-(N-K)} \right) .
\end{equation*}


The average probability of decoding failure at the destination is therefore equal to
$P_{\mathrm{f}} (N\!-\!K) \triangleq \mathbb{E} \left[ P_{\mathrm{f}} \left( N\!-\!K,E\right) \right]$,
where the expectation over $E$ depends implicitly on all possible channel realizations within a block.
While the average probability of decoding failure offers a good measure of performance, it alone does not capture the queueing behavior of the system.
Indeed, correlation among decoding-failure events may also alter the behavior of the queue at the transmitter.

\section{Arrival and Departure Processes}
\label{section:ArrivalsDepartures}

Having introduced a precise model for the physical layer, we turn to the description of the arrival and departure processes at the queue.
In our framework, the block-length, which we denote by $N$, remains fixed throughout and every codeword transmission requires $N$ consecutive uses of the channel.
Each data packet is broken into length-$K$ data segments that are separately encoded into blocks.
In terms of system characterization, $N$ is fundamental in that it determines the sampling period of our Markov chain.

We assume that the packet arrival process is i.i.d. Bernoulli with parameter $\gamma$.
This implies that, during each codeword transmission interval, a new packet arrives at the source with probability $\gamma$.
The number of bits in each data packet is assumed to be an i.i.d. random process whose marginal distribution is geometric with parameter $\rho$.
Therefore, the probability that a packet contains exactly $\ell$ bits becomes
\begin{equation*}
\Pr (L = \ell) = (1 - \rho)^{\ell - 1} \rho \qquad \ell = 1, 2, \ldots
\end{equation*}
where $\rho \in (0, 1)$.
These assumptions on the structure of the arrival process and the packet-length distribution are crucial for the construction of a tractable Markov model for our communication system.
They enable a rigorous analysis of the queue and lead to meaningful guidelines for system design and optimization.

Departures from the queue are governed by the underlying Gilbert-Elliot channel and the design-rate $r=K/N$ of our random linear code. 
The  number of information bits contained in every codeword is therefore $K = rN$.
A low-rate code will, in general, have a smaller probability of decoding failure than the same system with a higher rate code.
Still, the successful decoding of a codeword associated with a high-rate code leads to the transmission of a larger amount of data bits.
These competing considerations create a natural tradeoff between information content and probability of decoding failure.
Accordingly, the code-rate $r$, or equivalently the number of information bits $K$, is a parameter that should be optimized.

Once a code rate is selected, the number of successfully decoded codewords needed to complete the transmission of a given packet is $M = \left\lceil {L}/{rN} \right\rceil$.
Since $L$ is geometric, we find that $M$ also has a geometric distribution, albeit with parameter
\begin{equation*}
\rho_r = \sum_{\ell = 1}^{rN} (1 - \rho)^{\ell - 1} \rho
= 1 - (1 - \rho)^{rN} .
\end{equation*}
The probability that a data packet requires the successful transmission of $m$ data segments of size $rN$ is equal to
\begin{equation*}
\Pr (M = m) = \left( 1 - \rho_r \right)^{m-1} \rho_r \qquad m = 1, 2, \ldots
\end{equation*}
For a head packet to depart from the queue, the destination  must successfully decode the most recent codeword it received, and this codeword must carry the final segment of information corresponding to this packet.
Implicit to our system model is the ability of the destination to acknowledge the reception of a codeword through instantaneous feedback.
Based on this side information, the transmitter is able to remove data segments and packets from the queue after successful transmission.

\section{Queueing Behavior}
\label{section:QueueingBehavior}

The number of data packets in the queue at the onset of block~$s$ is denoted by $Q_s$.
The state of the Gilbert-Elliot channel at this same instant is represented by $C_{sN + 1}$.
Together, these two quantities form the state of our Markov process,
$U_s = \left( C_{sN+1}, Q_s \right)$.
We emphasize that the cardinality of this state space is countable, with $U_s$ belonging to $\{ b, g \} \times \mathbb{N}_0$.
Furthermore, the Markov chain underlying the evolution of our system possesses a special structure; it forms an instance of a discrete-time \emph{quasi-birth-death process}.
Fortunately, there are many established techniques to study such mathematical objects.
We present one possible approach in Section~\ref{section:MatrixGeometricMethod}.

The transition probability from $U_s$ to $U_{s+1}$ is given by
\begin{equation} \label{equation:StateTranstionProbabilities}
\begin{split}
\Pr ( U_{s+1} &= (d, q_{s+1}) | U_s = (c, q_s) ) \\
= \sum_{ e \in \mathbb{N}_0 }
&\Pr \left( Q_{s+1} = q_{s+1} | E=e, Q_s = q_s \right) \times \\
&\Pr \left( E=e, C_{(s+1)N+1} = d | C_{sN+1} = c \right) .
\end{split}
\end{equation}
Recall that a methodology was introduced in Section~\ref{section:ErasureDistribution} to derive the distribution of $\left( E, C_{(s+1)N+1} \right)$ conditioned on the value of $C_{sN+1}$.
Obtaining expressions for probabilities of the type
$\Pr \left( Q_{s+1} = q_{s+1} | E=e, Q_s = q_s \right)$
remains.

\begin{figure}[tbh!]
\begin{center}
\begin{psfrags}
\psfrag{e0}[c]{$(b,\!0)$}
\psfrag{e1}[c]{$(b,\!1)$}
\psfrag{e2}[c]{$(b,\!2)$}
\psfrag{g0}[c]{$(g,\!0)$}
\psfrag{g1}[c]{$(g,\!1)$}
\psfrag{g2}[c]{$(g,\!2)$}
\epsfig{file=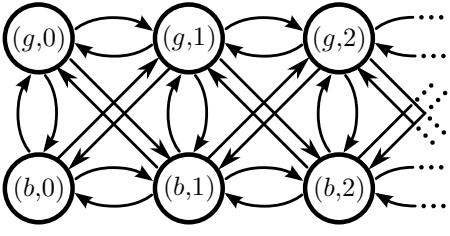,width=5.91cm}
\end{psfrags}
\caption{State space and transition diagram for the aggregate queued process $\{ U_s \}$;
self-transitions are intentionally omitted.}
\label{figure:QueuedSystem}
\end{center}
\end{figure}

We first consider conditional events $\{ Q_s = q_s \}$ for which $q_s > 0$; admissible values for $Q_{s+1}$ are then limited to values in $\{ q_s - 1, q_s, q_s + 1 \}$.
Two factors can affect the length of the queue, the arrival of a new data packet and the completion of a packet transmission.
The latter occurrence will only take place if a codeword is successfully decoded at the destination and the head packet has no additional data segment left at the source.
Keeping these facts in mind, we get
\begin{align*}
\begin{split}
\Pr & \left( Q_{s+1} = q_s +1 | E=e, Q_s = q_s \right) \\
&= \gamma \big( P_{\mathrm{f}}(N\!-\!K,e)
+ ( 1 - P_{\mathrm{f}}(N\!-\!K,e) ) (1 - \rho_r) \big)
\end{split} \\
\begin{split}
\Pr & \left( Q_{s+1} = q_s | E=e, Q_s = q_s \right)
= \gamma \left( 1 - P_{\mathrm{f}}(N\!-\!K,e) \right) \rho_r \\
&+ (1 - \gamma) \big( P_{\mathrm{f}}(N\!-\!K,e)
+ ( 1 - P_{\mathrm{f}}(N\!-\!K,e) ) (1 - \rho_r) \big)
\end{split} \\
\begin{split}
\Pr & ( Q_{s+1} = q_s - 1 | E=e, Q_s = q_s ) \\
&= (1 - \gamma) \left( 1 - P_{\mathrm{f}}(N\!-\!K,e) \right) \rho_r .
\end{split}
\end{align*}
When the queue is empty, $\{ Q_s = 0 \}$, only two possibilities can occur,
\begin{align*}
\Pr & ( Q_{s+1} = 1 | E=e, Q_s = 0 ) = \gamma \\
\Pr & (Q_{s+1} = 0 | E=e, Q_s = 0 ) = 1 - \gamma .
\end{align*}
Collecting these findings and using \eqref{equation:StateTranstionProbabilities}, we get the probability transition matrix of the Markov process $\{ U_s \}$.
A graphical rendition of the state transitions appears in Fig.~\ref{figure:QueuedSystem}.

To proceed with the analysis of our queued system, a compact representation of the conditional probabilities defined in \eqref{equation:StateTranstionProbabilities} is apropos.
For $q \in \mathbb{N}$ and $c, d \in \{ b, g \}$, we introduce the following mathematical notation,
\begin{align*}
\mu_{cd} &= \Pr ( U_{s+1} = (d, q-1) | U_s = (c, q) ) \\
\kappa_{cd} &= \Pr ( U_{s+1} = (d, q) | U_s = (c, q) ) \\
\lambda_{cd} &= \Pr ( U_{s+1} = (d, q+1) | U_s = (c, q) ) .
\end{align*}
Similarly, when the queue is empty, we use
$\kappa_{cd}^0 = \Pr ( U_{s+1} = (d, 0) | U_s = (c, 0) )$
and $\lambda_{cd}^0 = \Pr ( U_{s+1} = (d, 1) | U_s = (c, 0) )$.
Collectively, these labels define the 12 transition probabilities associated with a non-empty queue, and the 8 transition probabilities subject to the non-negativity constraint at zero.

%
%

We are ready to derive the equilibrium distribution of our system.
We note that, if the channel state is ergodic and the queue is stable, then the Markov chain $\{ U_s \}$ is positive recurrent and possesses a unique stationary distribution~\cite{Norris0521633966}.
Let $U = (C, Q)$ be a random vector with the following probability distribution,
\begin{equation*}
\Pr ( U = (c, q) ) = \lim_{s \rightarrow \infty} \Pr ( U_s = (c, q) ) .
\end{equation*}
We employ the semi-infinite vector $\pi$ as a convenient notation for the equilibrium distribution of our system, with
\begin{equation*}
\pi (2q + i) = \begin{cases} \Pr (C = b, Q = q) & \textrm{if }i = 1 \\
\Pr (C = g, Q = q) & \textrm{if }i = 2, \end{cases}
\end{equation*}
for $i\in \{1,2\}$ and $q\in \mathbb{N}_0$.
The states $\{ (b, q), (g, q) \}$ are known as the $q$th level of the Markov chain and $\pi_q \triangleq [\pi ( 2q + 1 ) \; \pi ( 2q + 2) ]$ is the stationary distribution associated with the $q$th level.

Using this compact notation, we can write the Chapman-Kolmogorov equations as
$\pi \mathbf{T} = \pi$,
where $\mathbf{T}$ is the probability transition matrix associated with $\{ U_s \}$.
One possible approach to solve for the stationary distribution of our Markov model is to employ spectral representation and ordinary generating functions~\cite{ParimalITW2010}.
In this article, we adopt an alternate means and apply the matrix geometric method~\cite{Neuts0486683427,Latouche0898714257}.

\subsection{Matrix Geometric Method}
\label{section:MatrixGeometricMethod}

We can represent the probability transition matrix $\mathbf{T}$ as a semi-infinite matrix of the form
\begin{equation} \label{equation:LinearNransformation}
\mathbf{T} = \left( \begin{array}{ccccc}
\mathbf{C}_1 & \mathbf{C}_0 & \mathbf{0} & \mathbf{0} & \cdots \\
\mathbf{A}_2 & \mathbf{A}_1 & \mathbf{A}_0 & \mathbf{0} & \cdots \\
\mathbf{0} & \mathbf{A}_2 & \mathbf{A}_1 & \mathbf{A}_0 & \cdots \\
\mathbf{0} & \mathbf{0} & \mathbf{A}_2 & \mathbf{A}_1 & \cdots \\
\vdots & \vdots & \vdots & \vdots & \ddots
\end{array} \right)
\end{equation}
where the submatrices $\mathbf{C}_1$, $\mathbf{C}_0$, $\mathbf{A}_2$, $\mathbf{A}_1$, and $\mathbf{A}_0$ are $2 \times 2$ real matrices.
More specifically, we have
\begin{xalignat*}{2}
\mathbf{A}_0 &= \left[ \begin{array}{cc} \lambda_{bb} & \lambda_{bg} \\
\lambda_{gb} & \lambda_{gg} \end{array} \right] &
\mathbf{A}_1 &= \left[ \begin{array}{cc} \kappa_{bb} & \kappa_{bg} \\
\kappa_{gb} & \kappa_{gg} \end{array} \right] \\
\mathbf{A}_2 &= \left[ \begin{array}{cc} \mu_{bb} & \mu_{bg} \\
\mu_{gb} & \mu_{gg} \end{array} \right] .
\end{xalignat*}
When the queue is empty, the relevant submatrices become
\begin{xalignat*}{2}
\mathbf{C}_0 &= \left[ \begin{array}{cc} \lambda_{bb}^0 & \lambda_{bg}^0 \\
\lambda_{gb}^0 & \lambda_{gg}^0 \end{array} \right] &
\mathbf{C}_1 &= \left[ \begin{array}{cc} \kappa_{bb}^0 & \kappa_{bg}^0 \\
\kappa_{gb}^0 & \kappa_{gg}^0 \end{array} \right] .
\end{xalignat*}
Note that the Markov chain associated with \eqref{equation:LinearNransformation} belongs to the class of processes with repetitive structure.
The following theorem characterizes its stationary distribution.

\begin{theorem}
Consider a positive recurrent Markov chain on a countable state space with transition matrix $\mathbf{T}$ given by \eqref{equation:LinearNransformation}.
Let the positive matrix $\mathbf{R}$ be defined as the limit, starting from $\mathbf{R}_0 = \mathbf{0}$, of the matrix recursion
\[ \mathbf{R}_{j+1} =  (\mathbf{A}_0 + \mathbf{R}_j^2 \mathbf{A}_2) (\mathbf{I} - \mathbf{A}_1)^{-1}. \]
Then, the $q$th-level stationary distribution $\pi_q$ satisfies $\pi_{q+1} = \pi_q \mathbf{R}$ for $q\geq 1$ with
%
%
$\pi_1 = \pi_0 \mathbf{Z}$ and
\begin{align*}
\mathbf{Z} &= (\mathbf{I} - \mathbf{C}_1) \mathbf{A}_2^{-1} \\
\pi_0 &= \left[ \begin{array}{cc} \frac{\beta}{\alpha+\beta} &
\frac{\alpha}{\alpha+\beta} \end{array} \right]
\left( \mathbf{I} + \mathbf{Z} (\mathbf{I} - \mathbf{R})^{-1} \right)^{-1} .
%
%
\end{align*}
\end{theorem}
\nocite{Neuts0486683427,Latouche0898714257}

\begin{corollary} \label{corollary:DecayRate}
The decay rate of the complementary cumulative distribution function of the queue satisfies
\begin{equation*} \label{equation:taildecay}
\lim_{\tau \rightarrow \infty} \tau^{-1} \log \Pr(Q \geq \tau)
= \log \varrho(\mathbf{R}) ,
\end{equation*}
where $\varrho(\mathbf{R})$ is the spectral radius of $\mathbf{R}$.
\end{corollary}

\section{Performance Evaluation}
\label{section:NumericalResults}


This mathematical characterization makes it possible to compute a wide range of advanced performance criteria for the system under consideration, including average packet error rate and outage capacity.
Herein, we focus on two measures that are most relevant to delay-sensitive communications.
First, we look at the probability that the queue exceeds a threshold, $\Pr (Q > \tau)$, where $\tau$ is relatively small.
Second, we examine the decay rate of the complementary cumulative distribution function, as discussed in Corollary~\ref{corollary:DecayRate}.
Again, we emphasize that the tail decay in buffer occupancy is given by the dominant eigenvalue of $\mathbf{R}$.

For illustrative purposes, we select the following parameters.
The Gilbert-Elliot erasure channel is defined by $\alpha = 0.02$, $\beta = 0.005$, $\varepsilon_b = 0.49$, and $\varepsilon_g = 0.0025$.
This generates an average erasure probability of $0.1$.
The channel memory decays at an exponential rate of $(1 - \alpha - \beta) = 0.975$.
The blocklength is fixed at $N=114$ and the arrival process is defined by the arrival probability $\gamma = 0.25$ and average packet length $\rho^{-1} = 195$.
If codewords are transmitted every 4.615~ms, then this corresponds to an arrival rate of roughly 10.6~Kbits/sec and an ergodic channel capacity of roughly 22.2~Kbits/sec.
These parameters are selected to loosely match the operation of a wireless GSM relay link.

\begin{figure}[t!]
\begin{center}
\begin{psfrags}
\psfrag{K}[c]{Information Bits per Block, $K$}
\psfrag{B}[c]{Tail Probability, $\Pr (Q > \tau)$}
\epsfig{file=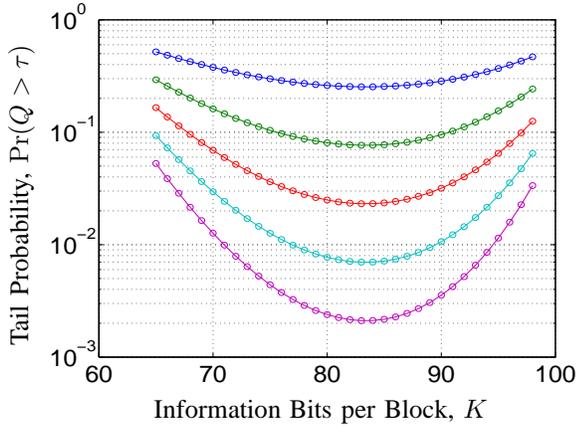,height=5.5cm,width=7.5cm}
\end{psfrags}
\caption{This figure shows tail probabilities in the equilibrium packet distribution of the queue, $\Pr (Q > \tau)$, for threshold values $\tau \in \{ 5, 10, 15, 20, 25 \}$.
The minimums occur uniformly at $rN=83$ for all threshold values.}
\label{figure:overflow1}
\end{center}
\end{figure}

System performance as a function of the number of information bits per codeword, $K$, is shown in Fig.~\ref{figure:overflow1}.
Each curve represents the complementary cumulative distribution function evaluated at a different threshold value, $\Pr (Q > \tau)$.

\begin{figure}[h!]
\begin{center}
\begin{psfrags}
\psfrag{A}[c][l][1][335]{Arrival Rate}
\psfrag{K}[c][r][1][15]{Information Bits $rN$}
\epsfig{file=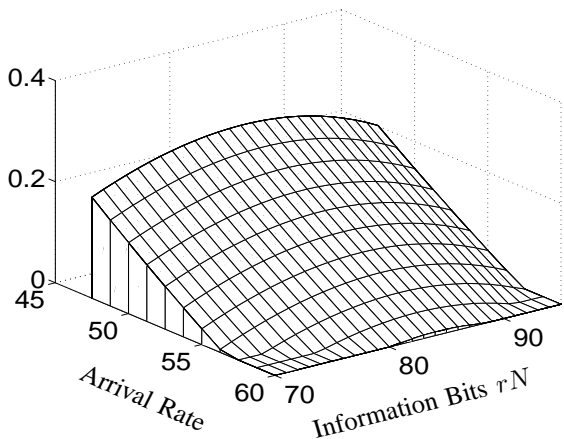,height=5.5cm,width=7.5cm}
\end{psfrags}
\caption{This figure shows tail decay rate, $-\lim_{\tau \rightarrow \infty} \tau^{-1} \log \Pr(Q \geq \tau)$, as a function of the number of information bits $rN$ and the average arrival rate ${\gamma}/{\rho_r}$ in bits per cycle.}
\label{figure:effcap1}
\end{center}
\end{figure}

As expected, the probability of the queue exceeding a prescribed threshold decreases as $\tau$ increases.
More interestingly, it is instructive to notice that $K=83$ appears uniformly optimal for all values of $\tau$.
Further supporting evidence for this observation is offered by looking at the asymptotic decay rate in tail occupancy, displayed in Fig.~\ref{figure:effcap1}.
When the arrival rate $\gamma \rho_r^{-1}$ is between 47.5 and 60, one finds that $K = 83$ is also optimal in terms of tail decay.
This robustness property is very encouraging, as it simplifies system design.

An important observation that does not appear on these two figures is the fact that, for short block lengths, the optimal value of $K$ depends heavily on the channel parameters $\alpha$, $\beta$, $\varepsilon_g$ and $\varepsilon_b$.
A naive conjecture would place $K = rN$ close to the Shannon limit $0.9 \times 114 = 102.6$, but this is much larger than the optimal value of $K=83$.
A more sophisticated approach is to maximize the throughput of a system with an infinite-backlog.
After some calculation, one finds that this leads to $K=87$, which is much closer to the true optimum.
But, as the channel memory parameter $(1 - \alpha - \beta)$ varies, the optimal value of $K$ changes substantially.
In fact, as $(1 - \alpha - \beta) \rightarrow 1$, $K$ approaches $N$.

\newpage

\section{Conclusions}
\label{section:Conclusion}

This work provides a unified approach that links queueing performance with the operation of a communication system at the physical layer.
The methodology and results are developed for the Gilbert-Elliot erasure channel, but can be generalized to more intricate finite-state channels with memory.
For example, the simple performance characterization of random codes over erasure channels extends naturally to hard-decision decoding of BCH codes over Gilbert-Elliot error channels.
For fixed parameters, the optimal code rate appears relatively insensitive to target threshold $\tau$ in the queue.
Still, channel memory and cross-over probabilities can affect this optimal operating point.
More generally, the optimal code rate seems to be linked to ratio between the codeword time and the coherence time of the channel.


\end{document}